\documentclass[twocolumn,amssymb, nobibnotes, showpacs, superscriptaddress, aps, prd]{revtex4}
\usepackage{amsmath}
\usepackage{epsfig}

\usepackage{graphicx}% Include figure files
\usepackage{dcolumn}% Align table columns on decimal point
\usepackage{bm}% bold math
\usepackage[bookmarks=true,
   colorlinks=true,
   linkcolor=blue,
   urlcolor=blue,
   citecolor=blue,
   bookmarks=true,
   hyperindex=true
]{hyperref}

\begin{document}

%Title of paper
\title{Calibration of atomic trajectories in a large-area dual-atom-interferometer gyroscope}

\author{Zhan-Wei Yao}
\affiliation{State Key Laboratory of Magnetic Resonance and Atomic and Molecular Physics, Wuhan Institute of Physics and Mathematics, Chinese Academy of Sciences, Wuhan 430071, China}
\affiliation{Center for Cold Atom Physics, Chinese Academy of Sciences, Wuhan 430071, China}
\author{Si-Bin Lu}
\affiliation{State Key Laboratory of Magnetic Resonance and Atomic and Molecular Physics, Wuhan Institute of Physics and Mathematics, Chinese Academy of Sciences, Wuhan 430071, China}
\affiliation{School of physics, University of Chinese Academy of Sciences, Beijing 100049, China}
\author{Run-Bing Li}
\email[]{rbli@wipm.ac.cn}
\affiliation{State Key Laboratory of Magnetic Resonance and Atomic and Molecular Physics, Wuhan Institute of Physics and Mathematics, Chinese Academy of Sciences, Wuhan 430071, China}
\affiliation{Center for Cold Atom Physics, Chinese Academy of Sciences, Wuhan 430071, China}
\author{Jun Luo}
\affiliation{State Key Laboratory of Magnetic Resonance and Atomic and Molecular Physics, Wuhan Institute of Physics and Mathematics, Chinese Academy of Sciences, Wuhan 430071, China}
\affiliation{Center for Cold Atom Physics, Chinese Academy of Sciences, Wuhan 430071, China}
\author{Jin Wang}
\email[]{wangjin@wipm.ac.cn}
\affiliation{State Key Laboratory of Magnetic Resonance and Atomic and Molecular Physics, Wuhan Institute of Physics and Mathematics, Chinese Academy of Sciences, Wuhan 430071, China}
\affiliation{Center for Cold Atom Physics, Chinese Academy of Sciences, Wuhan 430071, China}
\author{Ming-Sheng Zhan}
\email[]{mszhan@wipm.ac.cn}
\affiliation{State Key Laboratory of Magnetic Resonance and Atomic and Molecular Physics, Wuhan Institute of Physics and Mathematics, Chinese Academy of Sciences, Wuhan 430071, China}
\affiliation{Center for Cold Atom Physics, Chinese Academy of Sciences, Wuhan 430071, China}

\date{\today}

\begin{abstract}
% insert abstract here
We propose and demonstrate a method for calibrating atomic trajectories in a large-area dual-atom-interferometer gyroscope. The atom trajectories are monitored by modulating and delaying the Raman transition, and they are precisely calibrated by controlling the laser orientation and the bias magnetic field. To improve the immunity to the gravity effect and the common phase
noise, the symmetry and the overlapping of two large-area atomic interference loops are optimized by calibrating the atomic trajectories and by aligning the Raman-laser orientations. The dual-atom-interferometer gyroscope is applied in the measurement of the Earth rotation. The sensitivity is $1.2\times10^{-6}$ rad/s/$\sqrt{Hz}$, and the long-term stability is $6.2\times10^{-8}$ rad/s $@$ $2000$ s.
\end{abstract}

% insert suggested PACS numbers in braces on next line
% insert suggested keywords - APS authors don't need to do this
%\keywords{}

%\maketitle must follow title, authors, abstract, \pacs, and \keywords
\maketitle

\section{Introduction}

Gyroscopes have many important applications in scientific and technical fields\cite{Stedman1997a,Brosche1998a,Barbour1998a,Ciufolini2004a,Bernard2011a,Lefevre2013a}. As a novel rotation sensor, the atom-interferometer gyroscope presents an ultrahigh sensitivity\cite{Scully1993a,Gustavson2000a}. Since atom interferometers are both sensitive to the gravitational acceleration and the rotation, the gravity effect must be eliminated as more as possible in the rotation measurement. To cancel the gravity effect, the dual-atom-interferometer gyroscope has been designed and realized\cite{Canuel2006a,Gauguet2009a,Stockton2011a,Rakholia2014a}. The sensitivity of an atom interferometer can be improved by enlarging the atomic interference area and suppressing the phase noise. The atomic interference area was enlarged by increasing the interrogation times between two consecutive Raman pulses\cite{Tackmann2012a,Berg2015a}. The large-momentum-transfer beam splitters and mirrors were also developed by several groups, which is a potential technique for realizing a compact large-area atom-interferometer gyroscope. The Large-area dual-atom-interferometer gyroscope has important potential applications in the field of geophysics\cite{Barrett2014a,Dutta2016a}, and advances in the atom interferometer make it possible for a transportable atom-interferometer gyroscope\cite{Hauth2013a}.

\vskip 5pt
\noindent

However, the dual-atom-interferometer gyroscope is more sensitive to the gravity effect when the atomic interference area is enlarged. The symmetry and overlapping of atomic trajectories becomes very important to improve the performance of the dual-atom-interferometer gyroscope. If two atomic interference loops are not symmetrically overlapped, the gravity effect will cause a systematic error in the absolute rotation measurement. The phase noise, caused by the Raman lasers and the vibration, is also increased in a large-area atom interferometer. Thus, to improve the performance of the dual-atom-interferometer gyroscope, the symmetry and overlapping of two large-area atomic interference loops must be optimized by calibrating the atom trajectories.

\vskip 5pt
\noindent

In this paper, we demonstrate the calibration of atomic trajectories in a large-area dual-atom-interferometer gyroscope. The atomic trajectories are monitored by modulating and delaying the Raman transition, and they are precisely calibrated by controlling the laser orientation and the bias magnetic field. The wave-vector orientations of the Raman beams are precisely aligned by controlling the mirrors with the piezoelectric ceramics. After the wave-vector orientations are aligned and the atomic trajectories are well calibrated, the atom gyroscope is built based on two symmetric large-area atomic interference loops. The contrasts of two interference fringes are both larger than 20$\%$, and each interference area is 20 mm$^{2}$. The gravity induced phases have the same trends and they are canceled in two symmetric interference loops. The Earth rotation is measured by extracting the differential phase, where calibrating the atom trajectories is very important for the absolute rotation measurement.

\section{Experimental setup}

\begin{figure}[htp]
	\centering
		\includegraphics[width=0.42\textwidth]{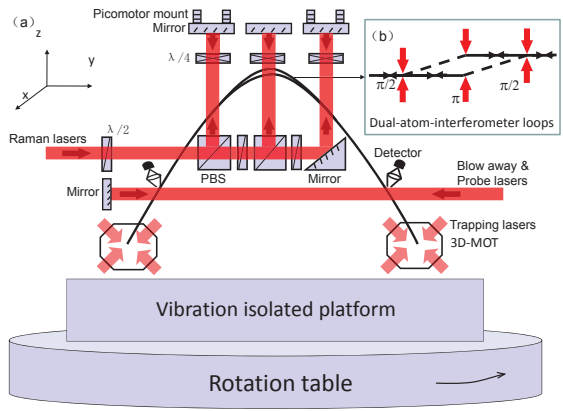}
	\caption{(color online) The schematic diagram of  experimental setup. An atom gyroscope is mounted on a vibration-isolated turntable. Three pairs of Raman lasers are counter-propagating along the gravity direction for measuring the horizontal rotation component and the symmetric dual-atom-interferometer loops are built by aligning the laser orientations and optimized by calibrating the atomic trajectories.}%
	\label{fig-1}%
\end{figure}

The schematic apparatus is shown in Fig.\textit{\ref{fig-1}}. The compact gyroscope is mounted on a vibration-isolated turntable. Similar to our previous work\cite{Yao2016a}, the $^{85}$Rb atoms are firstly loaded into two symmetric magneto-optical traps (MOTs) from the background vapor, and then they are launched by the moving optical molasses, with a velocity of 2.5 m/s and an angle of 14.0$^{\circ}$ with respect to the gravity direction. The polarization gradient cooling is applied when the atoms are accelerated from two MOTs. The cold atom cloud with the temperature of 5 $\mu$K are counter-propagating along the same parabolic trajectory. They are synchronously prepared to the initial state $|F=2,m=0\rangle$ by a microwave field and a blow-away laser. The initial atoms are manipulated by three pairs of separated Raman beams along the gravity direction. The residual magnetic field is compensated by three pairs of Helmholtz coils\cite{Li2008a}, and the ac Stark shift is canceled by adjusting the intensity ratio of Raman beams\cite{Li2009a}. The population of the state $|F=3,m=0\rangle$ is detected by the laser induced fluorescence, and the atomic interference fringes are observed by scanning the differential phase of Raman lasers.

The Raman lasers are prepared with an optical phase-locked loop\cite{Cao2017a}. The frequency difference of Raman lasers is 3.035 GHz, which corresponding to the clock transition of $^{85}$Rb atoms. The main laser is locked to the transition between $|5S_{1/2},F=3\rangle$ and $|5P_{3/2},F=4\rangle$ using the modulating transfer spectroscopy\cite{McCarron2008a}, and its frequency is red-shifted by an acoustic-optic modulator (AOM, 200 MHz) with the four-pass configuration\cite{Donley2005a}. The slave laser and the main laser are combined together with a polarization beam splitter (PBS), and the beat note signal is detected by a high-speed photo detector (Hamamatsu, G4176). To get the error signal and extend the frequency capture range, the beat note signal is detected by a phase frequency discriminator (Analog Device, AD9901). The phase error is feeded back to the current controller of the slave laser via a fast phase locking module (Toptica, mFALC). The frequency is chirped 2.5 MHz/s to compensate the Doppler shift induced by the gravity. To embrace more atoms participating interference, the two tapered amplifiers are used to increase the powers of the Raman lasers. The Raman lasers are switched by a 80-MHz AOM with a double-pass configuration. The intensity ratio of the Raman lasers is $1:2$ to cancel the ac Stark shift. The laser power after collimated into the fiber reaches 240 mW, which provides a Rabi frequency of 100 kHz. This corresponds the 50$\%$ atoms participating interference for the 5 $\mu$K temperature.

\begin{figure}[htp]
	\centering
		\includegraphics[width=0.42\textwidth]{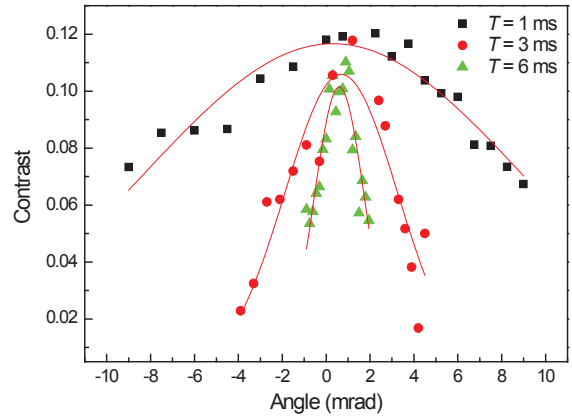}
	\caption{(color online) The dependence of the contrast of Ramsey-Bord\'{e} fringes on the orientations between two consecutive Raman beams. The contrast depends on the alignment of Raman beams as the interrogation time increased.}%
	\label{fig-2}%
\end{figure}

To build the larger atom interference loops, three pairs of separated counter-propagating Raman beams are applied along the gravity direction to manipulate the cold atoms as shown in Fig.\textit{\ref{fig-1}}. The Raman lasers are transmitted by a polarization maintaining fiber, and well collimated by an achromatic doublet lens with focal length of 150 mm. The Raman beams are split into three paths by using $\lambda /2$ wave plates, PBSs and mirrors. With the spacial separation of 3 cm, the interrogation time is $T$=52 ms between the two consecutive Raman pulses. To well overlap the atomic wave packets, three pairs of Raman beams should be precisely aligned at the level of several $\mu$rad \cite{Tackmann2012a}. In the experiment, two motor-driven mounts (New Focus, Picomotor) are used to adjust the mirrors. The mounts are remotely controlled by the piezo-electric ceramics drivers. First, the second pair of Raman beams are aligned along the gravity direction where the water surface is taken as a reference. The other two pairs of Raman beams are parallel to the second one. The crossing angles between the Raman beams and the gravity direction are less than several mrad. Second, the two symmetric Ramsey-Bord\'{e}(R-B) interferometers are used to adjust the orientations between two consecutive Raman beams. As shown in Fig.\textit{\ref{fig-2}}, the contrast of the R-B interference fringe depends on the alignment of Raman beams and the interrogation time. Third, the R-B interferometer is extended to embrace three pairs of Raman beams, and finally transform to the Mach-zehnder (M-Z) interferometer as the interrogation time increased. Thus, the atom packets well overlap, and the M-Z interference loops form when the three pairs of Raman beams are aligned and the three-pulse Raman sequence is applied.

\section{Calibration of atom trajectories}

The atom interferometer is sensitive to the rotation and the gravity, and the phase shift caused by the gravity and the rotation is written as

\begin{equation}\label{eq1}
\varphi=k_{eff} \cdot g T^{2}+2k_{eff} \cdot (\omega\times\nu) T^{2}+\phi_{laser}
\end{equation}
where, $k_{eff}$ is the effective wave vector of the Raman beams, $g$ is the gravitational acceleration, $T$ is interrogation time between two consecutive Raman pulses, $\omega$ is the rotation rate of the frame, $v$ is the velocity of atoms, and $\phi_{laser}$ is the phase of the Raman lasers. To measure the absolute rotation, the gravity effect should be eliminated as more as possible. In our experiment, two symmetric interference loops are built as shown in Fig.\textit{\ref{fig-1}} (b). Thanks to the inverse velocity, the rotation signal can be distinguished in dual atom interferometers by subtracting their phase shifts. The rotation induced phase shift, $\varphi$, can be differentially measured as $4k_{eff} \cdot (\omega\times\nu) T^{2}$ according to Eq. (\ref{eq1}). Furthermore, the common phase noise and the systematic error can be eliminated.

\begin{figure}[htp]
	\centering
		\includegraphics[width=0.43\textwidth]{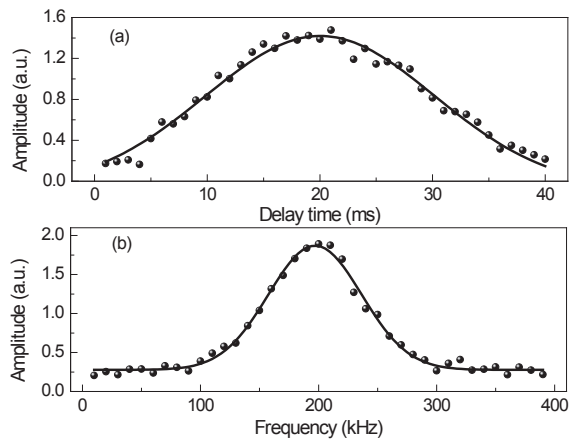}
	\caption{(color online) The atomic trajectory measurement using the interaction between the atoms and the Raman beams. (a) Delaying the Raman pulse interaction time to measure the horizontal position. (b) Scanning the Raman transition to measure the vertical velocity.}%
	\label{fig-3}%
\end{figure}

To decrease the systematic error and increase the contrast, the atomic trajectories of the dual atom interferometers should be perfectly symmetric and overlapped. The atomic trajectories are calibrated using the Raman transition. First, the atomic trajectories are monitored by the fluorescence signals of the Raman transition. As the Gaussian dependence of the atomic position on the Raman lasers, the atomic trajectories in different directions can be monitored by scanning the delay time and the two-photon detuning of the Raman lasers. Along the moving direction of the atoms, the atomic position is measured by controlling the delay time of the Raman pulse. By fitting the signal amplitude in Fig.\textit{\ref{fig-3}} (a) with the Gaussian function, the interaction time between the Raman lasers and the atom cloud is measured. In the vertical direction, we use the counter-propagating Raman transition to measure the atomic velocity. The Raman transition depends on the two-photon detuning as shown in Fig.\textit{\ref{fig-3}} (b). In the perpendicular direction to the atomic trajectory, the atomic position is measured by moving a slice which mounted on a motorized translation stage (Thorlabs, MTS25) behind the Raman beams. These results give a judging criterion for calibrating the two atomic trajectories. In the gravity field, the trajectory of an atom cloud depends on its initial position, velocity and orientation. During the moving molasses, the launch speed is controlled by the detuning of the cooling lasers, the launch angle is adjusted by the orientation of the cooling lasers, and  the initial position of atom cloud is shifted by adjusting the bias magnetic field. Finally, the atomic trajectories are calibrated by adjusting their launch speeds and directions, and optimized by monitoring the Raman transitions. After a series of modulations and measurements, two trajectories are overlapped in the horizontal and vertical directions and symmetric in the gravity direction. The position uncertainties in the horizontal and vertical directions are less than 0.08\% and  0.10\%, respectively. The atomic velocities along the gravity direction for each other are coincide with a uncertainty of 0.09\%. The crossover between the co-propagating and counter-propagating Raman transitions are avoided when the two atomic trajectories are perfectly optimized.

\begin{figure}[htp]
	\centering
		\includegraphics[width=0.42\textwidth]{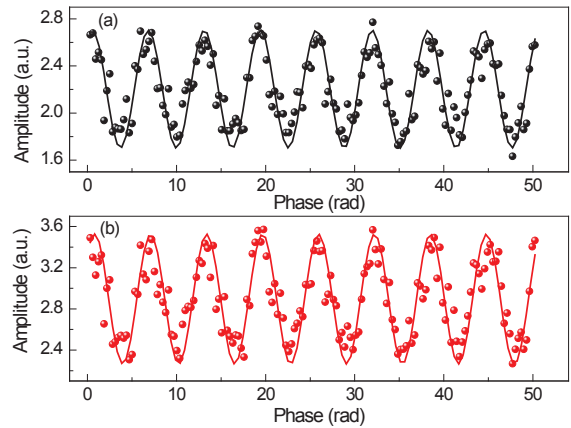}
	\caption{(color online)  The Mach-Zehnder fringes of the dual atom interferometers. The contrasts are 21.7$\%$ for the first interferometer and 22.7$\%$ for the second one, respectively.}%
	\label{fig-4}%
\end{figure}

By optimizing the symmetry of two atomic interference loops, the contrasts of atomic interference fringes are increased from 4$\%$ to more than 20$\%$. Fig.\textit{\ref{fig-4}}(a) shows the atomic interference fringes of the dual atom interferometers. Their contrasts are 21.7$\%$ and 22.7$\%$, respectively. The contrasts of two fringes are slightly different, which is mainly caused by the unbalance of three pairs of Raman beams. Although the vibration causes the phase noise in each atomic interference fringe, the differential phase noise can be suppressed in the dual atom interferometers.

\section{performance evaluation}

\begin{figure}[htp]
	\centering
		\includegraphics[width=0.42\textwidth]{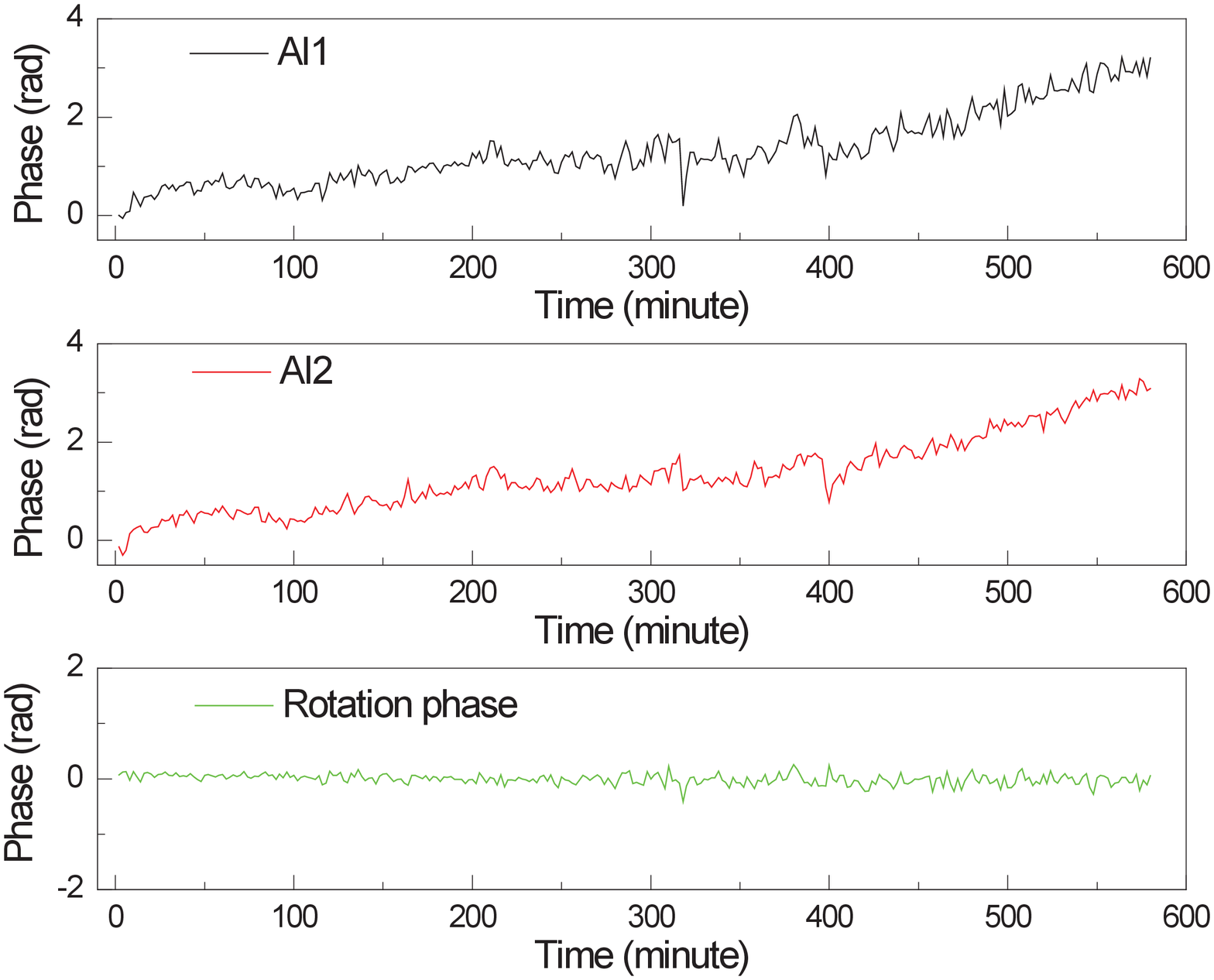}
	\caption{(color online)The dependence of the phase drift on the time. (a) The initial phases of the first atom interferometer. (b) The initial phases of the second atom interferometer. (c) The differential phase of the dual atom interferometers.}%
	\label{fig-5}%
\end{figure}

The orientations of the Raman beams are drifted with the temperature\cite{Durfee2006a}. In the dual-atom-interferometer gyroscope, the initial phases are extracted from atomic interference fringes, and they show a long-term drift caused by the temperature. The phases are seriously drifted as shown in Fig.\textit{\ref{fig-5}} (a) and (b). According to Eq.(\ref{eq1}), the rotation information is related to the differential phase between two atom interferometers as shown in Fig.\textit{\ref{fig-5}} (c). The drifts are canceled in the symmetric dual atom interferometers, thus the absolute rotation can be measured.

\begin{figure}[htp]
	\centering
		\includegraphics[width=0.42\textwidth]{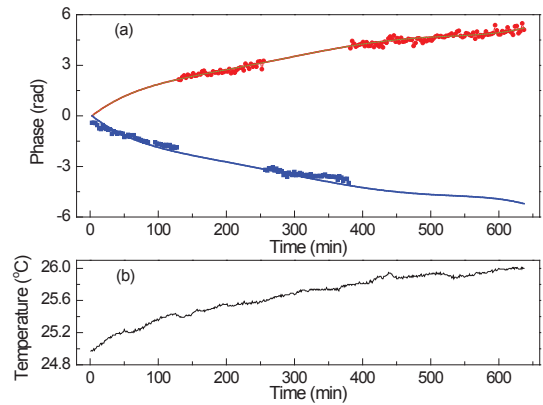}
	\caption{(color online) The initial phase and the temperature drift of the atom interferometers. (a) The initial phases of the first atom interferometer with (red dots) and without (blue squares) reversing the wave vector. (b)The temperature drift.  }%
	\label{fig-6}%
\end{figure}

Although this long-term drift can be canceled in the dual atom interferometers, it is necessary to suppress this drift in each atom interferometer\cite{cheinet2006a}. We synchronously monitor the temperature drift and the phase shifts of the dual atom interferometers. The phase shift of the first atom interferometer is shown in Fig.\textit{\ref{fig-6}} (a), and the temperature drift is shown in Fig.\textit{\ref{fig-6}} (b). The temperature drift causes a gravity-like phase shift in each atom interferometer. The phase is increased when the Raman lasers are propagating along the gravity (red dots), and decreased when the Raman wave vector is reversed (blue squares). The phase drift is inverse as the wave vector of the Raman lasers reversed, and the Pearson correlation coefficient between the phase shift and the temperature drift is 0.99. The wave-vector reversal method can eliminate the residual phase shift, including the wavefront, the Zeeman effect and the ac Stark effect.

\begin{figure}[htp]
	\centering
		\includegraphics[width=0.42\textwidth]{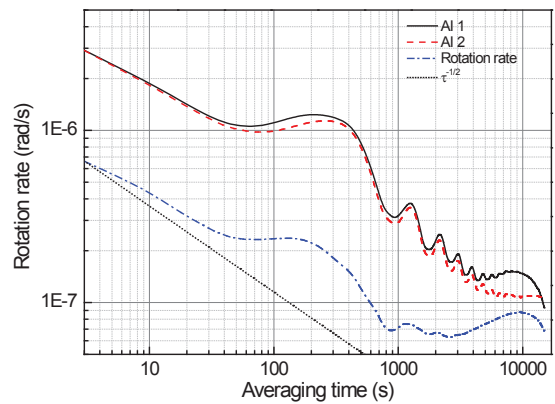}
	\caption{(color online)  The Allan deviations and the differential Allan deviation of the atom interferometers. The black solid line is for the first atom interferometer, and red dash line is  for the second one. The blue dot-dash line is the differential Allan deviation for the dual atom interferometers.}%
	\label{fig-7}%
\end{figure}

The performance of rotation measurement is evaluated by the Allan deviation\cite{Lawrence1996a}. To reduce the amplitude noise and shorten the sampling time, the square wave modulation is used to analyze the data\cite{cheinet2006a}. Interference signal is set to two adjacent mid-fringe points by adjusting the phase difference of the Raman lasers. The populations, $P(\psi)$ and $P(\psi+\pi)$, are measured, and the common noise is suppressed by subtracting $P(\psi)$ from $P(\psi+\pi)$. The vibration noise and the laser phase noise are differentially rejected in the dual atom interferometers. To resolve the gravity drift, a feedback loop is applied, where the added signals $P(\psi)$ + $P(\psi+\pi)$ are timely feedback to the Raman lasers. The minus signals $P(\psi)$ - $P(\psi+\pi)$ are the rotation information. Fig.\textit{\ref{fig-7}} shows the Allan deviations for the first atom interferometer (black solid line) and for the second one (red dash line).

The temperature induced phase drift is also compensated in both two atom interferometers by the feedback loop. The differential Allan deviation between two atom interferometers is also shown in Fig.\textit{\ref{fig-7}} (blue dot-dash line), where the white noise is better suppressed. The humps in the individual Allan deviation curves are caused by the temperature oscillations, and they are suppressed in the differentially extracted gyroscope signal curve by using the feedback loop. The sensitivity of the atom gyroscope is $1.2\times10^{-6}$ rad/s/$\sqrt{Hz}$, and the long-term stability is $6.2\times10^{-8}$ rad/s over the time of 2000 s.

\section{absolute rotation measurement}

\begin{figure}[htp]
	\centering
		\includegraphics[width=0.45\textwidth]{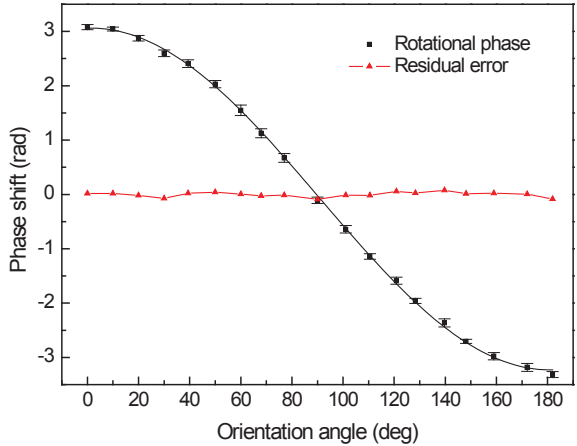}
	\caption{(color online) The measurements of the Earth's rotation rate by modulating the turntable. The rotation induced phase shift depends on the sensor orientation angle in laboratory coordinates, and it is a sinusoidal function.}%
	\label{fig-8}%
\end{figure}

The absolute rotation measurement will have important applications in inertial navigation, precision measurement and in geophysics. To measure the absolute Earth's rotation rate, the experimental setup is mounted on an air-bearing turntable, which is driven by a direct-torque motor. The rotation angle of the turntable is calibrated by a precision goniometer (Renishaw RESM). The atomic interference area is along the horizontal direction. The projection of the Earth rotation on the atomic interference area are changed as the setup rotated. The Earth rotation induced phase shift is the differential phase of the dual atom interferometers, which is a sinusoidal function as shown in Fig.\textit{\ref{fig-8}} (black squares). The Raman lasers with and without reversing wave vectors are used to improve the accuracy of the rotation measurement.

\vskip 5pt
\noindent

The Earth's rotation rate is extracted by fitting the curve (black line), and it is ($6.24\pm0.03$)$\times10^{-5}$ rad/s. Considering the latitude in our location, the Earth' rotation rate is ($7.32\pm0.03$)$\times10^{-5}$ rad/s, which is consistent with the value provided by the International Earth Rotation Service (IERS). The residual errors between the experimental data and the fitting results are shown in Fig.\textit{\ref{fig-8}} (red triangles). These errors are mainly caused by the magnetic field gradient, which causes the unbalanced phase rejections in the dual atom interferometers. The uncertainty of the Earth's rotation rate is $0.5\%$  with the value predicted. This result provides a potential candidate for seeking north orientation of the Earth.

\vskip 10pt
\noindent

\section{conclusion}

In summary, the symmetry and overlapping of atomic trajectories are very important to improve the performance of the dual-atom-interferometer gyroscope. In this work, we demonstrate a method for calibrating the atomic trajectories in the large-area dual-atom-interferometer gyroscope. The large-area dual-atom-interferometer loops are built after the relative orientations of Raman mirrors are precisely aligned with the piezo-electric ceramics. The atomic trajectories are monitored by the Raman transitions and precisely calibrated by controlling the laser orientation and the bias magnetic field. The symmetry and overlapping of two atom-interferometer loops are improved after the atomic trajectories are precisely calibrated. The common phase noise is suppressed, the gravity effect is canceled, and the temperature drift is compensated.  The short-term sensitivity of atom gyroscope is $1.2\times10^{-6}$ rad/s/$\sqrt{Hz}$, the long-term stability is $6.2\times10^{-8}$ rad/s after an integrating time of 2000 s. The performance is mainly limited by the amplitude noise, which can be improved by the normalized detection. The Earth's rotation rate is measured by modulating a turntable, and it is agrees with the value predicted in the IERS. This work is helpful to develop the high-precision gyroscope, which can be used to monitor the Earth's rotation rate and seek its north orientation, and to develop the inertial instrument.

\vskip 5pt
\noindent

We acknowledge the financial support from the National Key Research and Development Program of China under Grant No. 2016YFA0302002, the National Natural Science Foundation of China under Grant Nos. 91536221, 11674362, 91736311, the Strategic Priority Research Program of Chinese Academy of Sciences under Grant No. XDB21010100, and the Youth Innovation Promotion Association of Chinese Academy of Sciences.

\bibliography{basename of .bib file}

\end{document}